\UseRawInputEncoding

\documentclass[journal,10pt]{IEEEtran}

\usepackage{color}
  \usepackage[dvips]{graphicx}

\usepackage[cmex10]{amsmath}
\usepackage{amssymb}
\DeclareMathOperator*{\argmin}{\arg\!\min}
\begin{document}

\title{Amplifier-Coupled Tone Reservation for Minimization of OFDM Nonlinear Distortion}

\author{Pawel Kryszkiewicz,~\IEEEmembership{Member,~IEEE}
\thanks{Copyright (c) 2018 IEEE. Personal use is permitted. For any other purposes, permission must be obtained from the IEEE by emailing pubs-permissions@ieee.org.
This is the author's version of an article that has been published in this journal. Changes were made to this version by the publisher prior to publication.
The final version of record is available at http://dx.doi.org/10.1109/TVT.2018.2795339}%
\thanks{P. Kryszkiewicz is with the Chair of Wireless Communications, Poznan University of Technology, Poland e-mail: pawel.kryszkiewicz@put.poznan.pl}
\thanks{The work presented in this paper has been funded by
the National Science Centre in Poland within the EcoNets
project based on decision no. DEC-2013/11/B/ST7/01168 and by the Polish Ministry of Science and Higher Education funds for the status activity
project no. 08/81/8124.}}

\markboth{IEEE Transactions on Vehicular Technology,~Vol.~XX, No.~XX, XXX~2017}
{}

\maketitle

\begin{abstract}
Nonlinear distortion of an OFDM signal is a serious problem when it comes to energy-efficient Power Amplifier (PA) utilization. Typically, Peak-to-Average Power Ratio (PAPR) reduction algorithms and digital predistortion algorithms are used independently to fight the same phenomenon. This paper proposes an Amplifier-Coupled Tone Reservation (AC-TR) algorithm for the reduction of nonlinear distortion power, utilizing knowledge on the predistorted PA characteristic. The optimization problem is defined. Its convexity is proved. A computationally-efficient solution is presented. Finally, its performance is compared against two state-of-the-art TR algorithms by means of simulations and measurements. The results show the proposed solution is advantageous, both in terms of nonlinear distortion power and the required number of computations. 
\end{abstract}


%
\IEEEpeerreviewmaketitle

\section{Introduction}
%
%
%
%
As the 5G technology is approaching, attention is drawn to the energy-efficiency of wireless systems \cite{Cavalcante_energy_efficient_5G}. High influence on energy consumption is exerted by power amplifiers (PAs) in mobile terminals and base stations \cite{Auer_energy_cons_LTE_2011}. Additionally, power amplifiers should provide high bandwidth, dynamic range and introduce a small amount of distortions. However, the accomplishment of these requirements depends on the applied modulation scheme. 

Nowadays, Orthogonal Frequency Division Multiplexing (OFDM) is typically used. Unfortunately, it has high variations of transmitted signal amplitude. These variations are commonly measured by means of Peak-to-Average Power Ratio (PAPR). 
In order to amplify such a signal without introducing distortion an amplifier of a linear gain up to maximal power sample is needed, while the mean transmitted signal power is much lower. Such a PA will be highly energy inefficient. Another possibility is to allow distortions introduced to the peak samples, while obtaining high mean transmit power and high energy efficiency. While a 5G system can possibly use a more advanced multicarrier technique, e.g., Filter Bank Multicarrier \cite{FBMC_2011} or Non-Orthogonal Frequency Division Multiplexing \cite{Bogucka_Radar_Sonar_2011} the nonlinear distortion power will be even higher. It has been observed that the utilization of a pulse shape different than a rectangular window (used in OFDM), will typically increase the PAPR metric\cite{Chafii_2016_PAPR_optimal_OFDM}. 

There are two main techniques used in order to obtain high energy efficiency of OFDM transmission while maintaining low nonlinear distortion power. First, the nonlinearity characteristic of a power amplifier can be linearized, e.g., by means of digital predistortion (DPD) \cite{Fu_predis_freq_selective_2015,Kryszkiewicz_ISWCS_2015_predistortion}. The predistorter model has to be adjusted to the PA model. Contemporary solid-state power amplifiers can be characterized accurately using, e.g., Volterra Series or Three-Box Model \cite{Gharaibeh_book_nonlinear}. The more effects are to be described by the model, e.g., the memory effect, the more complicated it becomes. The Rapp model \cite{Gharaibeh_book_nonlinear} is a relatively simple one, introducing only amplitude-amplitude distortion. However, as it is able to describe the most significant part of nonlinear distortion, its modification has been recently proposed as a reference model by 3GPP \cite{Nokia_3gpp_Rapp}. In the case of perfect predistortion, the effective nonlinearity (joint characteristic of the predistorter and PA) has a soft-limiter-like characteristic, i.e., the output signal is not distorted up to some level and clipped above this level\footnote{Soft-limiter-like characteristic is the ultimate limit requiring perfect PA characterization and unlimited complexity of DPD.}. It means that all signal samples exceeding a given level contribute to nonlinear distortion. A common approach to reduce nonlinear distortion power even further is PAPR reduction \cite{Bogucka_PAPR_overview}.
 One of the most promising PAPR reduction methods is the Tone Reservation (TR) technique that has already been introduced in the DVB-T2 standard  \cite{ETSI_DVB_T2}. A subset of all modulated subcarriers is devoted to special symbols adjusted in order to minimize the peak power. There are various approaches to optimizing these values. The optimal solution for peak-power reduction has been proposed in \cite{Tellado_TR_1998} (referred here as PAPR-TR) and followed by suboptimal algorithms, e.g., \cite{Hussain_gradient_TR_2009, Jiang_CF_TR_2014}.  However, all of them focus on the minimization of peak power instead of minimization of total nonlinear distortion power. It is possible that the TR technique will minimize the maximal peak, but the total power of the clipped signal will be increased (because of the introduction of many smaller peaks). Recently, new solutions have appeared that try to reduce nonlinear distortion using TR. In \cite{Deumal_Cumic_minimization_2011}, the cubic metric has been minimized. Although it makes the result closer to distortion power minimization, it does not take into account the PA characteristic. Similarly, in \cite{Hu_TR_Nonlinearity_2015}, a non-optimal solution has been proposed in order to maximize the correlated signal at the output of the soft-limiter amplifier. An optimal TR solution for the minimization of power of all clipped samples for a soft-limiter PA has been proposed in \cite{Kryszkiewicz_TR_EW_2017}. In \cite{Gazor_2012}, the authors propose a TR scheme maximizing signal to distortion ratio assuming soft-limiter amplifier. Unfortunately, this algorithm not only modifies the symbols modulating reserved tones but data symbols as well, i.e., it is more like precoding. However, all these solutions are not tightly connected with the utilized effective (e.g., after predistortion) PA model.

In this paper, the nonlinear distortion at the output of a Rapp-modeled amplifier is minimized using the proposed Amplifier-Coupled TR (AC-TR) method. Thanks to a smoothness parameter '$p$', this model can be adjusted both for highly nonlinear amplifiers, e.g., $p\approx 2$, and soft-limiter-like amplifiers, e.g., $p>10$. Additionally, the clipping threshold of the amplifier can be adjusted, giving the AC-TR method the most degrees of freedom in modeling the PA characteristic among all published TR solutions. The assumption of a PA characteristic known at the OFDM modulator is realistic, as predistortion is nowadays commonly implemented in the digital domain. The proposed optimization function is proved to be convex, which allows for finding globally optimal TR values efficiently using Newton's method. The computational complexity of the solution is significantly reduced thanks to the use of Fast Fourier Transform (FFT) blocks.  Although the utilization of a more advanced PA model might be advantageous in terms of resultant distortion power, the computational complexity will increase and the convexity property might be lost.     

This paper is organized as follows. Section II defines the considered system with the presentation of a classical TR method shown in Section II A, and the proposed AC-TR method shown in Section II B. The simulation results, measurement outcomes and computational complexity evaluation are shown in Section III. The paper is concluded in Section IV.               

\section{System Model}
\begin{figure}[!t]
\centering
\includegraphics[width=3.5in]{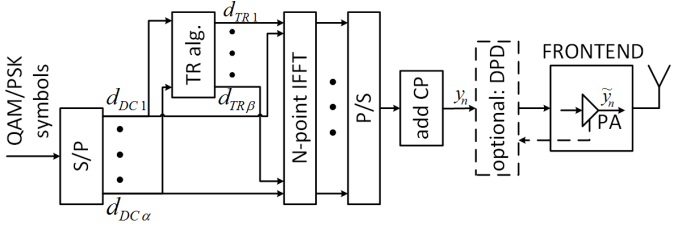}
\caption{OFDM transmitter diagram.}
\label{fig_model}
\end{figure}

The considered system model is depicted in Fig. \ref{fig_model}. It is an OFDM modulator based on the N-point Inverse Fast Fourier Transform (IFFT) block utilizing $\alpha$ data subcarriers and $\beta$ TR subcarriers. S\textbackslash P and P\textbackslash S denote serial-to-parallel and parallel-to-serial conversion, respectively. Assuming all subcarriers at the input of IFFT are indexed in the range $\{ -N/2,..., N/2-1\}$, the vectors of data and TR subcarriers indices are $\mathbf{D}=\{D_u\}$ and $\mathbf{T}=\{T_l\}$ where $u=1,...,\alpha$ and $l=1,...,\beta$ \footnote{Typically the $0$-th subcarrier is not modulated to meet frontend limitations.}. For a single OFDM symbol preceded by $N_{\mathrm{CP}}$ samples of the cyclic prefix (CP), modulated with vector $\mathbf{d_{DC}}$ of QAM/PSK data symbols and vertical vector $\mathbf{d_{TR}}$ of TR symbols the time-domain signal is
\begin{equation}
y_{n}=\frac{1}{\sqrt{N}}\sum_{u=1}^{\alpha}d_{\mathrm{DC}u}e^{\imath 2 \pi \frac{nD_u}{N}}
+
\frac{1}{\sqrt{N}}\sum_{l=1}^{\beta}d_{\mathrm{TR}l}e^{\imath 2 \pi \frac{nT_l}{N}}
\label{eq_time_domain}
\end{equation}     
for $n=-N_{\mathrm{CP}},...,N-1$. The above formula can be simplified by assuming that $x_n=\frac{1}{\sqrt{N}}\sum_{u=1}^{\alpha}d_{\mathrm{DC}u}e^{\imath 2 \pi \frac{nD_u}{N}}$ and $F_{n,l}=\frac{1}{\sqrt{N}}e^{-\imath 2 \pi \frac{nT_l}{N}}$ giving
\begin{equation}
y_{n}=x_n+\sum_{l=1}^{\beta}d_{\mathrm{TR}l}F^{*}_{n,l},
\label{eq_time_domain_simplified}
\end{equation}
where $^{*}$ denotes the complex conjugate.
The signal at the output of Rapp-modeled effective (i.e., possibly after DPD) power amplifier is \cite{Gharaibeh_book_nonlinear}:
\begin{equation}
\tilde{y}_{n}=\frac{Gy_{n}}{\left(1+\frac{|y_{n}|^{2p}}{V^{2p}}\right)^{\frac{1}{2p}}}
\label{eq_Rapp_output}
\end{equation}     
where $p$ is smoothing factor, $G$ represents the linear gain and $V$ is the input saturation amplitude. Without the loss of generality, the unity linear gain ($G=1$) can be considered. Observe that for high values of $p$ this model approaches the soft-limiter/clipper. 

Any $\tilde{y}_{n}$ can be factorized into a scaled input signal $\lambda y_{n}$ and uncorrelated distortion signal $s_{n}$ as

\begin{equation}
\tilde{y}_{n}=\lambda y_{n} + s_{n},
\label{eq_bussgang}
\end{equation}   
where the correlation coefficient is
  \begin{equation}
\lambda=\frac{\mathbb{E}\left[\tilde{y}_{n}y^{*}_{n}\right]}
{\mathbb{E}\left[y_{n}y^{*}_{n}\right]},
\label{eq_corr_coef}
\end{equation}  
and $\mathbb{E}$ denotes expectation. This is the background for the homogenous linear mean square estimation presented in \cite{papoulis2002probability}. 

 Assuming no TR algorithm is applied, i.e., $y_n=x_n$, the OFDM signal is assumed to be complex-Gaussian distributed \cite{Wei_2010_rozklad_OFDM}. Therefore,$|y_n|=r_n$ is Rayleigh distributed and the mean transmit power equals $\mathbb{E}\left[|y_{n}|^2\right]=\sigma^{2}$. In this case the correlation coefficient can be calculated for the Rapp amplifier as
  \begin{align}
\lambda=\frac{1}{\sigma^2}\int_{0}^{\infty}\frac{r_{n}}{\left(1+\frac{r_{n}^{2p}}{V^{2p}}\right)^{\frac{1}{2p}}} 
r_n 2\frac{r_n}{\sigma^2}e^{-\frac{r_{n}^{2}}{\sigma^{2}}}\,dr_n. 
\label{eq_calka_lambda_0}
\end{align} 
It can be simplified by defining the Input Back-Off as a ratio of the maximum input unclipped signal power (for $p\rightarrow \infty$) to the mean input signal power, i.e., $\mathrm{IBO}=\frac{V^{2}}{\sigma^2}$. 
Although many different IBO definitions are proposed, the same formulation as in \cite{Tavares_IBO_def_2016} is used here.
Utilizing additional substitution of $\xi=\frac{r_{n}}{\sigma}$, equation (\ref{eq_calka_lambda_0}) simplifies to
  \begin{align}
\lambda=\int_{0}^{\infty}\frac{2\xi^{3}}{\left(1+\left(\frac{\xi^{2}}{\mathrm{IBO}}\right)^{p}\right)^{\frac{1}{2p}}}
e^{-\xi^{2}}\,d\xi
\label{eq_calka_lambda_1}
\end{align} 
which can be integrated numerically.
As it will be shown in Sec. III, $\lambda$ increases with $p$ and IBO with its maximum equals 1 when $\tilde{y}_{n} =y_{n}$ and $s_n=0$.

In this paper, Signal-to-Distortion Ratio (SDR) is mainly used to evaluate TR methods. It is defined as a ratio of the mean correlated power at the output of PA to the mean distortion power, i.e.,
  \begin{align}
\mathrm{SDR}=\frac{|\lambda|^2 
\sum_{j=1}^{\alpha}\mathbb{E}\left[ |d_{\mathrm{DC}j}|^{2} \right]
}{N \mathbb{E}\left[|s_{n}|^{2}\right]}.
\label{eq_SDR_def}
\end{align} 
Observe that signal power is calculated only for data symbols. The SDR allows one to indirectly estimate OFDM reception performance. The increase in TR symbols power is irrelevant to the reception quality. 
Most importantly, the utilized definition is a wideband SDR, i.e., covering both in-band and out-of-band frequencies. Although per-subcarrier SDR \cite{Araujo_nonlinear_OFDM_2010} would be more informative, it would make comparison of TR algorithms more difficult.
\subsection{PAPR minimizing TR method}
 The classic TR optimization method is defined as \cite{Tellado_TR_1998}
\begin{equation}
\mathbf{\widehat{d_{\mathrm{TR}}}}=\argmin_{\mathbf{d_{\mathrm{TR}}}}
\max_{n}\left|x_{n}+\sum_{l=1}^{\beta}d_{\mathrm{TR}l}F^{*}_{n,l}\right|,
\label{eq_problem_standard_TR}
\end{equation}
i.e., the maximum amplitude out of $y_n$ samples defined in (\ref{eq_time_domain_simplified}) is minimized by adjusting complex symbols in vector $\mathbf{d_{\mathrm{TR}}}$. It is typically solved by means of Second Order Cone Programming using a software package like CVX\cite{cvx}. Recently, an optimized TR solution has been proposed in \cite{Aggarwal_TR_IP_method_2006}. It uses the interior point method in order to solve this problem, while reducing computational complexity. However, it is not a pure TR method, as in addition to the calculation of reserved tones, data symbols are modified. However, straightforward modifications can be applied in order to use this method to solve problem (\ref{eq_problem_standard_TR}).

\subsection{Proposed AC-TR method}
The optimization problem is defined as
\begin{equation}
\min_{\mathbf{d_{\mathrm{TR}}}}
f(\mathbf{d_{\mathrm{TR}}}),
\end{equation}
where the optimization function is
\begin{align}
f(\mathbf{d_{\mathrm{TR}}})&=
\sum_{n=-N_{\mathrm{CP}}}^{N-1}
|\tilde{y}_{n}-K y_{n}
|^{2}
\end{align}
and $K$ is a positive real value being the optimization parameter. Substituting $\lambda y_{n} + s_{n}$ for $\tilde{y}_{n}$ it is obtained $|(\lambda-K)y_{n}+ s_{n}|^{2}$. Knowing that $y_{n}$ is orthogonal to $s_{n}$ for $\lambda$ calculated according to (\ref{eq_corr_coef}), the expectation over the optimization function can be calculated giving 
\begin{align}
\mathbb{E}\left[ f(\mathbf{d_{\mathrm{TR}}})\right]
=
\sum_{n=-N_{\mathrm{CP}}}^{N-1}
(K-\lambda)^{2} \sigma^{2}
+\mathbb{E}\left[|s_{n}|^{2}\right].
\label{eq_exp_funkcja_celu}
\end{align}
It can be shown that $\lambda\in\langle 0;1\rangle$ for any input signal and amplifier parameters. One approach will be to choose $K=\lambda$. This will cause the minimization of average uncorrelated distortion power $\mathbb{E}\left[|s_{n}|^{2}\right]$. However, neither is $\lambda$ known prior to transmission, nor can the problem be guaranteed to be convex. According to Appendix \ref{sec_appendix_convex}, the problem is convex for $K\geq1$. In this case uncorrelated nonlinear distortion power $\mathbb{E}\left[|s_{n}|^{2}\right]$ should be reduced, $\lambda$ should be increased and TR symbols power should be limited (being part of $\sigma^{2}$) according to (\ref{eq_exp_funkcja_celu}). However, too high $K$ value will put most effort on minimization of the $K^2\sigma^{2}$ component, i.e., the reduction of TR symbols power. A balance is obtained by choosing $K=1$. Additionally, this $K$ value changes the optimization into a minimization of the squared error between signals on the input and output of the power amplifier.  
Although $K=1$ is the parameter of choice, all derivations in this section will be done in a general case.  

 The optimization function can be simplified by utilizing the repeatability of CP samples and (\ref{eq_Rapp_output}), giving
\begin{align}
f(\mathbf{d_{\mathrm{TR}}})&=
\sum_{n=0}^{N-1-N_{\mathrm{CP}}}
\left|
y_{n}\right|^{2}\left(
b_n
^{-\frac{1}{2p}}\!\!-\!\!K\right)^{2}
\\ \nonumber &
+
2\!\sum_{n=N-N_{\mathrm{CP}}}^{N-1}
\left|
y_{n}\right|^{2}\left(
b_n
^{-\frac{1}{2p}}\!\!-\!\!K\right)^{2},
\end{align}
where $y_n$ is defined in (\ref{eq_time_domain_simplified}) and $b_n=1+V^{-2p}|y_n|^{2p}$.
It can be observed that
\begin{equation}
|y_n|^{2}=\left(\Re y_n\right)^{2}+\left(\Im y_n\right)^{2},
\end{equation}
where $\Re x$ and $\Im x$ are the real and imaginary parts of a complex number $x$ and
\begin{equation}
\Re y_n=\Re x_n+\sum_{l=1}^{\beta}\Re F_{n,l}\Re d_{\mathrm{TR} l}+
\Im F_{n,l}\Im d_{\mathrm{TR} l},
\end{equation}
\begin{equation}
\Im y_n=\Im x_n+\sum_{l=1}^{\beta}\Re F_{n,l}\Im d_{\mathrm{TR} l}-\Im F_{n,l}\Re d_{\mathrm{TR} l}.
\end{equation}
The optimal TR symbols are obtained for all $2\beta$ first-order partial derivatives equal to zero, i.e., 
\begin{equation}
\mathbf{d_{\mathrm{TR}}}:
\begin{cases}
\forall_{l=1,...,\beta}
\frac{\partial f
}{
\partial \Re d_{\mathrm{TR} l}
}
=0
\\
\forall_{l=1,...,\beta}
\frac{\partial f
}{
\partial \Im d_{\mathrm{TR} l}
}=0.
\end{cases}
\end{equation}
\subsubsection{Problem solution}
The optimization problem can be solved efficiently using Newton's method. Let us denote $\mathbf{d_{\mathrm{TR}}^{(\mathrm{k})}}$ as $k$-th approximation of TR symbols obtained as
\begin{equation}
\label{eq_newton_step}
\left[\!\!
\begin{array}{c}
\Re \mathbf{d_{\mathrm{TR}}^{(\mathrm{k})}}\!
\\
\Im \mathbf{d_{\mathrm{TR}}^{(\mathrm{k})}}\!
\end{array}\!
\right]\!=\!\left[\!\!
\begin{array}{c}
\Re \mathbf{d_{\mathrm{TR}}^{(\mathrm{k-1})}}\!
\\
\Im \mathbf{d_{\mathrm{TR}}^{(\mathrm{k-1})}}\!
\end{array}\!
\right]\!\!
-\!\!\left(\bigtriangledown^{2}
f\left(\mathbf{d_{\mathrm{TR}}^{(\mathrm{k-1})}}\right)
\right)^{-1}
\!\!\!\!
\bigtriangledown f\left(\mathbf{d_{\mathrm{TR}}^{(\mathrm{k-1})}}\right),
\end{equation}
where $\mathbf{X}^{-1}$ denotes the inverse of matrix $\mathbf{X}$, $\bigtriangledown^{2}
f\left(\mathbf{d_{\mathrm{TR}}^{(\mathrm{k-1})}}\right)$ and 
$\bigtriangledown
f\left(\mathbf{d_{\mathrm{TR}}^{(\mathrm{k-1})}}\right)$ is a $2\beta \times 2\beta$ real Hessian matrix and a $2\beta \times 1$ real Jacobian vector calculated for function $f\left(\mathbf{d_{\mathrm{TR}}}\right)$ at $\mathbf{d_{\mathrm{TR}}^{(\mathrm{k-1})}}$, respectively.
The Jacobian is defined by two $\beta \times 1$ vectors as
\begin{equation}
\bigtriangledown f\left(\mathbf{d_{\mathrm{TR}}^{(\mathrm{k})}}\right)
=\left[
\begin{array}{c}
\mathbf{q}^{(k)}\\\mathbf{w}^{(k)} 
\end{array}
\right],
\end{equation}
where
$q^{(k)}_l=\frac{\partial f}{
\partial \Re d_{\mathrm{TR} l} }|_{\mathbf{d_{\mathrm{TR}}^{(\mathrm{k})}}}$ and 
$w^{(k)}_l=\frac{\partial f}{
\partial \Im d_{\mathrm{TR} l} }|_{\mathbf{d_{\mathrm{TR}}^{(\mathrm{k})}}}$.
The Hessian is defined by four $\beta \times \beta$ submatrices as
\begin{equation}
\bigtriangledown^{2}
f\left(\mathbf{d_{\mathrm{TR}}^{(\mathrm{k})}}\right)=\left[ \begin{array}{c c}
\mathbf{A}^{(k)}
&
\mathbf{B}^{(k)}
\\
\mathbf{C}^{(k)}
&
\mathbf{D}^{(k)}
\end{array}
\right],
\end{equation}
where 
$A^{(k)}_{l,m}=\frac{\partial^{2} f
}{
\partial \Re d_{\mathrm{TR} l} \partial \Re d_{\mathrm{TR} m}
}|_{\mathbf{d_{\mathrm{TR}}^{(\mathrm{k})}}}$,
 $B^{(k)}_{l,m}=\frac{\partial^{2} f
}{
\partial \Re d_{\mathrm{TR} l} \partial \Im d_{\mathrm{TR} m}
}|_{\mathbf{d_{\mathrm{TR}}^{(\mathrm{k})}}}$, 
$C^{(k)}_{l,m}=\frac{\partial^{2} f
}{
\partial \Im d_{\mathrm{TR} l} \partial \Re d_{\mathrm{TR} m}
}|_{\mathbf{d_{\mathrm{TR}}^{(\mathrm{k})}}}$ and $D^{(k)}_{l,m}=\frac{\partial^{2} f
}{
\partial \Im d_{\mathrm{TR} l} \partial \Im d_{\mathrm{TR} m}
}|_{\mathbf{d_{\mathrm{TR}}^{(\mathrm{k})}}}$. 
The respective partial derivatives are calculated as
\begin{equation}
q^{(k)}_l=\sum_{n=0}^{N-1} \Gamma^{(k)}_{n} \left( 
\Re y_n^{(k)} \Re F_{n,l}-\Im y_n^{(k)} \Im F_{n,l}
\right)
\label{eq_q_l_def}
\end{equation} 
\begin{equation}
w^{(k)}_l=\sum_{n=0}^{N-1}\Gamma^{(k)}_{n} \left( 
\Re y_n^{(k)} \Im F_{n,l}+\Im y_n^{(k)} \Re F_{n,l}
\right),
\end{equation} 
\begin{align}
&A^{(k)}_{l,m}\!=\!\!\sum_{n=0}^{N-1}\!\!\Gamma^{(k)}_{n} \left( 
\Re F_{n,m}\Re F_{n,l}+\Im F_{n,m} \Im F_{n,l}
\right)
\!+\!\Lambda^{(k)}_{n} 
\\& \nonumber
\cdot\!\left(\!\Re y_n^{(k)} \Re F_{n,l}\!-\!\Im y_n^{(k)} \Im F_{n,l}\!\right)
\!\!\left(\! \Re y_n^{(k)} \Re F_{n,m}\!-\!\Im y_n^{(k)} \Im F_{n,m}\!\right)\!
,
\end{align} 
\begin{align}
\label{eq_B}
&B^{(k)}_{l,m}\!=\!\!\sum_{n=0}^{N-1}\!\!\Gamma^{(k)}_{n} \left( 
\Im F_{n,m}\Re F_{n,l}-\Re F_{n,m} \Im F_{n,l}
\right)
\!+\!\Lambda^{(k)}_{n} 
\\& \nonumber
\cdot\!\left(\!\Re y_n^{(k)} \Re F_{n,l}\!-\!\Im y_n^{(k)} \Im F_{n,l}\!\right)
\!\!\left(\!\Re y_n^{(k)} \Im F_{n,m}\!+\!\Im y_n^{(k)} \Re F_{n,m}\!\right)\!
,
\end{align}
\begin{align}
&C^{(k)}_{l,m}\!=B^{(k)}_{m,l},
\end{align}  
\begin{align}
\label{eq_D_l_m}
&D^{(k)}_{l,m}\!=\!\!\sum_{n=0}^{N-1}\!\!\Gamma^{(k)}_{n} \left( 
\Re F_{n,m}\Re F_{n,l}+\Im F_{n,m} \Im F_{n,l}
\right)
\!+\!\Lambda^{(k)}_{n} 
\\& \nonumber
\cdot\!\left(\!\Re y_n^{(k)} \Im F_{n,l}\!+\!\Im y_n^{(k)} \Re F_{n,l}\!\right)
\!\!\left(\!\Re y_n^{(k)} \Im F_{n,m}\!+\!\Im y_n^{(k)} \Re F_{n,m}\!\right)\!
,
\end{align} 
where $\mathbf{\Gamma}^{(k)}$ and $\mathbf{\Lambda}^{(k)}$ are real value vectors defined 
as
\begin{equation}
\Gamma^{(k)}_{n}\!\!\!=\!\!\!
\begin{cases}
\!\!2\!\!\left(\!\!K\!\!-\!\!b_n^{(k)-\frac{1}{2p}}\!\right)\!\!\left(\!K\!\!-\!\!b_n^{(k)-\frac{1}{2p}-1}\!\right)\! & \!\!\!\!\text{if}~n\!\!\in\!\! \{0,...,\!N\!\!-\!\!1\!\!-\!\!N_{\mathrm{CP}}\!\} \\
\!\!4\!\!\left(\!\!K\!\!-\!\!b_n^{(k)-\frac{1}{2p}}\!\right)\!\!\left(\!K\!\!-\!\!b_n^{(k)-\frac{1}{2p}-1}\!\right)\! & \!\!\!\!\text{if}~n\!\!\in\!\! \{N\!\!-\!\!N_{\mathrm{CP}},...,\!\!N\!\!-\!\!1\!\},
\end{cases}
\end{equation}
\begin{equation}
\Lambda^{(k)}_{n}\!\!\!=\!\!\!
\begin{cases}
\!\!\frac{2}{V^{2p}}\!|y_n^{(k)}\!|^{2p-2}
b_n^{(k)-\frac{1}{2p}-1}
\!\!\left(\!\!K\!\!-\!\!b_n^{(k)-\frac{1}{2p}-1} \right.
\\
\left.\!\!\!+\!\left(\!1\!\!+\!\!2p\right)b_n^{(k)-1}\!\!\!
\left(\!\!K\!\!-\!\!b_n^{(k)-\frac{1}{2p}}\!\right)
\!\right)
 & \!\!\!\!
\!\!\!\!\!\!\!\!\!\!\!\!\!\!\!\!\!\!\!\!\!\!
\text{if}~n\!\!\in\!\! \{0,...,\!N\!\!-\!\!1\!\!-\!\!N_{\mathrm{CP}}\!\} \\
\!\!\frac{4}{V^{2p}}\!|y_n^{(k)}\!|^{2p-2}
b_n^{(k)-\frac{1}{2p}-1}
\!\!\left(\!\!K\!\!-\!\!b_n^{(k)-\frac{1}{2p}-1} \right.
\\
\left.\!\!\!+\!\left(\!1\!\!+\!\!2p\right)b_n^{(k)-1}\!\!\!
\left(\!\!K\!\!-\!\!b_n^{(k)-\frac{1}{2p}}\!\right)
\!\right)
 & \!\!\!\!
\!\!\!\!\!\!\!\!\!\!\!\!\!\!\!\!\!\!\!\!\!\!\!
\text{if}~n\!\!\in\!\! \{N\!\!-\!\!N_{\mathrm{CP}},...,\!N\!\!-\!\!1\!\}.
\end{cases}
\end{equation}

The computational complexity of the above calculation can be reduced by calculating $\mathbf{\Gamma}^{(k)}$ and $\mathbf{\Lambda}^{(k)}$ only once in a given iteration. Additionally, the properties of DFT and its fast computation algorithms can be utilized. Let us denote DFT of $N$-length vector $\mathbf{X}$ evaluated at frequency bin $m$ (addressed cyclically within set $\{-N/2,..., N/2-1\}$) as
\begin{equation}
\mathcal{F}(\mathbf{X})_m=\frac{1}{\sqrt{N}}\sum_{n=0}^{N-1}X_n e^{-\imath 2 \pi \frac{n m}{N}},
\label{eq_DFT_fast}
\end{equation}
and element-wise vectors multiplication as $\odot$.
With a modicum of algebra applied to (\ref{eq_q_l_def})-(\ref{eq_D_l_m}), we obtain
\begin{equation}
q^{(k)}_l+\imath w^{(k)}_l=\mathcal{F}(\mathbf{\Gamma}^{(k)}\odot\mathbf{y}^{(k)})_{T_l},
\label{eq_Jakobian_fast}
\end{equation}
\begin{align}
A^{(k)}_{l,m}\!\!&=
\frac{0.5}{\sqrt{N}}\Re \mathcal{F}(\mathbf{\Lambda}^{(k)}\!\!\odot\!\!\mathbf{y}^{(k) 2})_{T_l+T_m}
\nonumber \\ &
\!\!+\!\!\frac{1}{\sqrt{N}}\Re \mathcal{F}(
\mathbf{\Gamma}^{(k)}+
0.5\mathbf{\Lambda}^{(k)}\!\!\odot\!\!|\mathbf{y}^{(k)}|^{2})_{T_l-T_m},
\label{eq_szybkie_A}
\end{align}
\begin{align}
B^{(k)}_{l,m}\!\!&=
\frac{0.5}{\sqrt{N}}\Im \mathcal{F}(\mathbf{\Lambda}^{(k)}\!\!\odot\!\!\mathbf{y}^{(k) 2})_{T_l+T_m}
\nonumber \\ &
\!\!-\!\!\frac{1}{\sqrt{N}}\Im \mathcal{F}(
\mathbf{\Gamma}^{(k)}+
0.5\mathbf{\Lambda}^{(k)}\!\!\odot\!\!|\mathbf{y}^{(k)}|^{2})_{T_l-T_m},
\label{eq_szybkie_B}
\end{align}
\begin{align}
D^{(k)}_{l,m}\!\!&=
-\frac{0.5}{\sqrt{N}}\Re \mathcal{F}(\mathbf{\Lambda}^{(k)}\!\!\odot\!\!\mathbf{y}^{(k) 2})_{T_l+T_m}
\nonumber \\ &
\!\!+\!\!\frac{1}{\sqrt{N}}\Re \mathcal{F}(
\mathbf{\Gamma}^{(k)}+
0.5\mathbf{\Lambda}^{(k)}\!\!\odot\!\!|\mathbf{y}^{(k)}|^{2})_{T_l-T_m}.
\label{eq_szybkie_D}
\end{align}
\subsubsection{Computational complexity}

The computational complexity of the proposed AC-TR algorithm can be estimated by calculating the number of arithmetical operations needed to solve (\ref{eq_newton_step}) and obtain time-domain OFDM symbol samples $y_n^{(k)}$ in $k$-th iteration. An operation is defined here as a single real-numbers addition, subtraction, multiplication, division, power or comparison. 

First, $|y_n^{(k-1)}|^{2}$ is calculated that is followed by the calculation of $b_n^{(k-1)}$, $\Gamma_n^{(k-1)}$ and $\Lambda_n^{(k-1)}$ requiring about $20N$ operations. Next, the Jacobian of function $f(\mathbf{d}_{\mathrm{TR}})$ is calculated according to (\ref{eq_Jakobian_fast}) by $N$ complex-by-real numbers multiplications, i.e., $2N$ operations, and an $N$-point complex number FFT. Assuming the utilization of a split-radix algorithm \cite{Duhamel_split_radix_1986}, it requires $4N\log_2 N-6N+8$ operations. The Hessian matrix creation starts with the calculation of FFT of $\mathbf{\Lambda}^{(k)}\odot\mathbf{y}^{(k) 2}$ and $\mathbf{\Gamma}^{(k)}+0.5\mathbf{\Lambda}^{(k)}\odot|\mathbf{y}^{(k)}|^{2}$. Knowing that the second vector has only real values and that a real-number FFT requires about half the number of operations \cite{Duhamel_split_radix_1986}, it requires $6N\log_2 N+N+12$ operations. 
Hessian $\bigtriangledown^{2} f\left(\mathbf{d_{\mathrm{TR}}^{(\mathrm{k-1})}}\right)$ of a convex function $f(\mathbf{d_{\mathrm{TR}}})$ is a positive semidefinite matrix according to \cite{boyd2004convex}. As such a submatrix $\mathbf{C}^{(k-1)}$ is obtained from $\mathbf{B}^{(k-1)}$ by means of symmetry. Similarly, submatrices $\mathbf{A}^{(k-1)}$ and $\mathbf{D}^{(k-1)}$ are symmetrical, so that only $\frac{\beta}{2}(\beta+1)$ entries have to be calculated requiring $2\frac{\beta}{2}(\beta+1)$ operations according to (\ref{eq_szybkie_A}) and (\ref{eq_szybkie_D}). 
According to (\ref{eq_szybkie_B}), $\beta^{2}$ operations have to be carried to calculate $\mathbf{B}^{(k-1)}$. 
Finally, based on a positive semidefinite Hessian matrix, (\ref{eq_newton_step}) can be solved efficiently using the Cholesky decomposition, and forward/backward substitutions of triangular matrices.
The Cholesky decomposition of a $2\beta \times 2 \beta$ Hessian matrix is invoked, requiring $\frac{8}{3} \beta^{3}+2\beta^{2}+\frac{1}{3}\beta$ operations \cite{Hunger}. It is followed by forward and backward substitution in order to solve linear system (\ref{eq_newton_step}) requiring a total of $8\beta^2$ operations. The iteration is finalized by calculating the IFFT of TR symbol updates and adding the resultant complex vector to vector $\mathbf{y}^{(k-1)}$, giving $\mathbf{y}^{(k)}$. In total, a single AC-TR iteration requires $14N\log_2 N+13N+\frac{8}{3}\beta^{3}+12\beta^{2}+\frac{4}{3}\beta+28$ operations. 

Similarly, the numbers of operations have been calculated for two competitive TR algorithms explained and used in the next section, i.e., the PAPR-TR \cite{Aggarwal_TR_IP_method_2006} and NCC-TR \cite{Hu_TR_Nonlinearity_2015} algorithms. In the case of PAPR-TR, the main computational cost comes from 4 $N$-point (I)FFT calculations (one of them is real-valued). Additionally, a positive semidefinite system of $2\beta$ equations has to be solved using the Cholesky decomposition and forward/backward substitution (as in the AC-TR algorithm). In total, it results in $14N\log_2 N+10N+\frac{8}{3}\beta^{3}+26\beta^{2}+\frac{1}{3}\beta+28$ operations per iteration. The computational complexity of the NCC-TR algorithm is dominated by 2 $N$-point, complex-numbers (I)FFTs per iteration. Additionally, as the algorithm operates only on time domain samples exceeding clipping threshold $V$, its computational complexity depends on the number of these samples (denoted here by $\Theta$). The total number of operations per iteration equals $8N\log_2 N+14\Theta+13$. These results are summarized in Table \ref{table_complexity}. As the total computational complexity per OFDM symbol depends on the required number of iterations and the number of clipped samples $\Theta$, the numerical comparison will be shown in the next section.
\begin{table}[]
\centering
\caption{Computational complexity of TR algorithm per iteration}
\label{table_complexity}
\begin{tabular}{|l|l|}
\hline
\textbf{Algorithm} & \textbf{Number of arithmetical operations per iteration} \\ \hline
AC-TR             										& $14N\log_2 N+13N+\frac{8}{3}\beta^{3}+12\beta^{2}+\frac{4}{3}\beta+28$ \\ \hline
PAPR-TR \cite{Tellado_TR_1998,Aggarwal_TR_IP_method_2006}        & $14N\log_2 N+10N+\frac{8}{3}\beta^{3}+26\beta^{2}+\frac{1}{3}\beta+28$ \\ \hline
NCC-TR \cite{Hu_TR_Nonlinearity_2015} & $8N\log_2 N+14\Theta+13$                 \\ \hline
\end{tabular}
\end{table}      

\section{Simulation Results}
The proposed AC-TR method is compared in this section with the classical TR method minimizing PAPR (PAPR-TR) \cite{Tellado_TR_1998} and a TR method maximizing the normalized correlation coefficient (NCC-TR) \cite{Hu_TR_Nonlinearity_2015}. PAPR-TR is implemented using an optimal, yet computationally efficient method based on \cite{Aggarwal_TR_IP_method_2006}. However, differently from \cite{Aggarwal_TR_IP_method_2006}, the PAPR-TR method does not modify the symbols on data subcarriers. The OFDM modulator utilizes an IFFT of $N=1024$ points and a cyclic prefix of $N_{\mathrm{CP}}=N/8$ samples. The total set of occupied subcarrier indices is $\{-100,...,-1\}\bigcup \{1,...,100\}$. It means that about 4 times oversampling is implemented. This results from the fact that typically, PA characterization and digital predistortion is carried out after signal oversampling \cite{Boumaiza_PA_char_2007}\footnote{Although, the AC-TR method works effectively without oversampling as well.}. There are $\beta=11$ reserved subcarriers of arbitrarily \footnote{Although the number and location of reserved tones can change the absolute values of utilized metrics, e.g., SDR, it is not expected to change significantly the performance of TR algorithms in relation to each other.} chosen indices 
$\mathbf{T}=\{-100,-80,-60,-40,-20,-1,20,40,60,80,100\}$, used for TR symbols or null symbols, when one of the TR algorithms is used, or the reference system (no TR algorithm) is considered, respectively. The other subcarriers are modulated with random QPSK symbols. Simulations not shown here confirmed that similar results are obtained for other QAM/PSK constellations. While the PAPR-TR algorithm operates independently from the PA model, NCC-TR assumes a soft-limiter PA with a known clipping threshold $V$. The proposed AC-TR utilizes both clipping threshold $V$ and smoothing factor $p$. As stated previously, $K$ is equal to 1, as it guarantees convexity and both reduces the nonlinear distortion power and increases the correlation coefficient $\lambda_{\mathrm{TR}}$. In each case, $10^4$ random OFDM symbols are generated. Most importantly, for fair comparison, the same clipping threshold $V$ is used for all systems, based on the IBO value for the reference system, i.e., $x_n=y_n$. This is denoted as \emph{ref. IBO} in the figures below. As such, the data symbols should have the same transmit power in all the cases, no matter what the power on reserved subcarriers is. The results were obtained for $p=4$ (modeling a typical, nonlinear PA) and $p=10$ (modeling the PA after digital predistortion) if not stated otherwise. As all considered TR methods are iterative, a common stopping condition is chosen for fair comparison. A given TR algorithm stops at the $k$-th iteration if the maximal absolute change of the TR symbol is lower than 0.01, i.e., $\max_l \left|d_{\mathrm{TR}l}^{(k)}-d_{\mathrm{TR}l}^{(k-1)}\right|<0.01$. It is reasonable, as $\mathbb{E}[|d_{\mathrm{DC} u}|^{2}]=1$. Initial simulations have shown no significant change in achievable SDR for higher accuracy of TR symbol calculation.  

\begin{figure}[!t]
\centering
\includegraphics[width=3.5in]{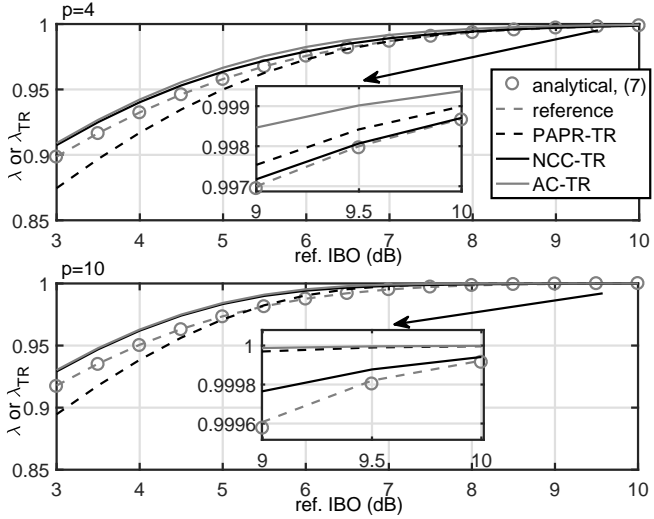}
\caption{Correlation coefficient $\lambda$ for all considered systems while varying reference IBO.}
\label{fig_lambda_vs_IBO}
\end{figure}

First, for all considered systems, the signal at the output of PA was divided into a correlated term (individually calculated correlation parameter according to (\ref{eq_corr_coef})) and uncorrelated nonlinear distortion. The correlation coefficients are shown in Fig. \ref{fig_lambda_vs_IBO}. It is visible that analytical solution (\ref{eq_calka_lambda_1}) overlaps with the results for the reference system. After the application of any TR method, signal $y_n$ is not a complex-Gaussian one, so that (\ref{eq_calka_lambda_1}) is not valid. The application of the AC-TR method provides the strongest wanted signal (highest $\lambda$) at the output of the amplifier in all the cases. Most interestingly, for relatively low IBO values, the PAPR-TR method is outperformed even by the reference system. The PAPR-TR method minimizes the strongest signal peak, possibly introducing many smaller peaks exceeding the clipping threshold. Therefore, higher nonlinear distortion power and a smaller correlation coefficient is expected.       

\begin{figure}[!t]
\centering
\includegraphics[width=3.5in]{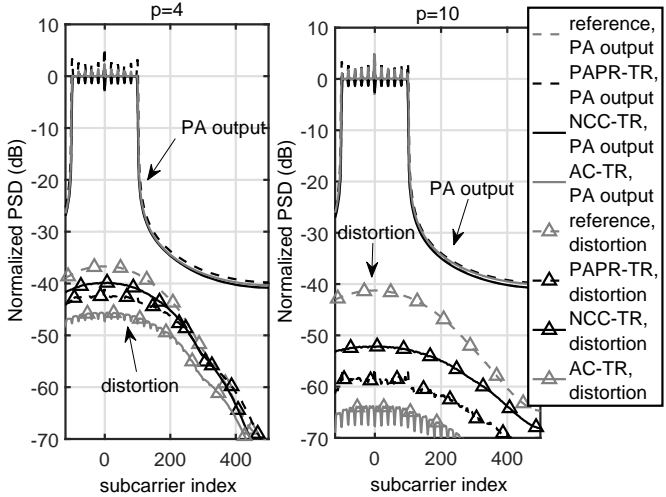}
\caption{Normalized PSDs for reference IBO=8 dB and $p$ equal 4 or 10.}
\label{fig_psd}
\end{figure}

In Fig. \ref{fig_psd}, normalized Power Spectra Densities (PSDs) are shown when reference IBO=8 dB for signals at the output of the PA and their uncorrelated distortion parts. The chosen IBO value is relatively high, so that all signals PSDs at the PA output are overlapping, i.e., the effect of the OFDM subcarrier spectrum sidelobes \cite{Kryszkiewicz_OCCS_2013} dominates over nonlinear distortion. However, focusing on nonlinear distortion signals only (plots with markers), it is visible that AC-TR outperforms all other solutions, both in-band and out-of-band of the wanted signal. Notably, for a more linear PA ($p=10$), the obtained nonlinear distortion power is lower for all considered systems in comparison to the PA of $p=4$.     

\begin{figure}[!t]
\centering
\includegraphics[width=3.5in]{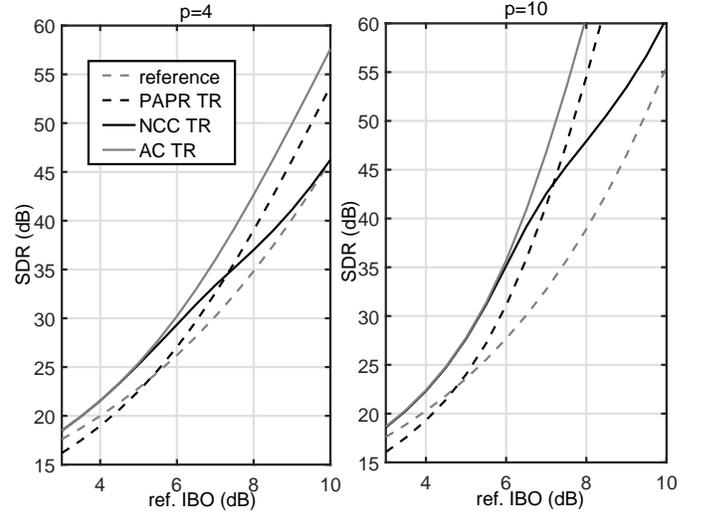}
\caption{SDR as a function of reference IBO for all considered systems and $p$ equal 4 or 10.}
\label{fig_SDR_IBO}
\end{figure}

In Fig. \ref{fig_SDR_IBO}, SDR defined as in (\ref{eq_SDR_def}) is calculated for all considered systems for two $p$ values and various reference IBO values. It is visible that the AC-TR method guarantees the maximal SDR value in all cases. For IBO equal to 7 dB and $p$ equal to 10, the gain over NCC-TR, PAPR-TR and the reference system is 4.3 dB, 5.5 dB and 14.1 dB, respectively. Observe that predistortion itself (change from $p=4$ to $p=10$ for the reference system) provides only a 2.5 dB gain. On the other hand, the utilization of the AC-TR method in a non-predistorted system ($p=4$) provides an about 7.5 dB gain in comparison to the reference system. The combination of both techniques achieves the synergy effect, gaining about 15 dB in terms of SDR (change from the reference system for $p=4$ to AC-TR for $p=10$). While the NCC-TR method results in SDR close to the results of the AC-TR method for lower IBO values, it is outperformed by two other TR algorithms for higher IBO values. On the other hand, the PAPR-TR method results in dropped SDR values in comparison to the reference system for low IBO values. This algorithm can introduce signal amplitude peaks, increasing nonlinear distortion power, while maintaining the minimal value of the maximal signal amplitude peak.    

\begin{figure}[!t]
\centering
\includegraphics[width=3.5in]{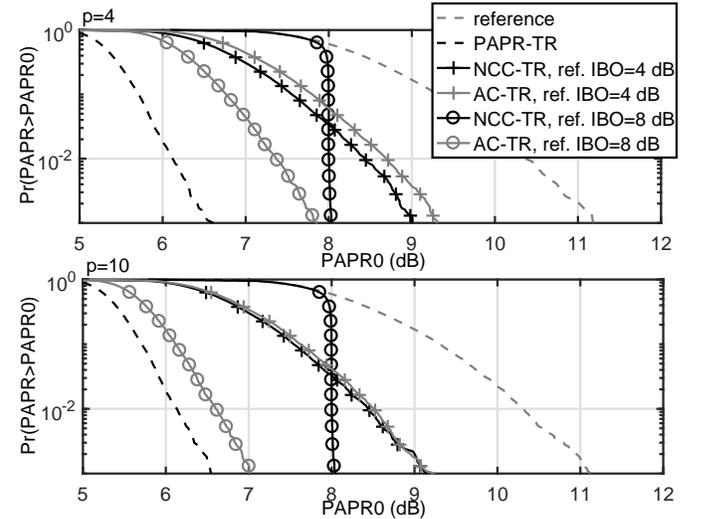}
\caption{PAPR CCDFs for reference IBO=4 dB or reference IBO=8 dB and $p$ equal 4 or 10}
\label{fig_PAPR}
\end{figure}

Finally, Complementary Cumulative Distribution Functions (CCDFs) of the PAPR metric are plotted in Fig. \ref{fig_PAPR}. Observe that while the PAPR-TR method guarantees a minimal PAPR, the reference system is characterized by the highest probability of high PAPR values. PAPR distribution is independent from the PA characteristic, i.e., $p$ and IBO, for the reference and PAPR-TR systems. Most interestingly, PAPR distribution for both $p$ and reference IBO equal to 4 dB is similar for both the AC-TR and NCC-TR methods. There are so many peaks contributing to nonlinear distortion that it can be more beneficial to keep a high maximal amplitude while reducing the power of smaller amplitude peaks. This is the reason for relatively high PAPR of NCC-TR and AC-TR algorithms in this case. The difference between both $p$ values is visible only for the AC-TR method. In the case of $p$ equal to 10 and relatively high IBO (IBO=8 dB), the main part of nonlinear distortion comes from clipping the maximal peak. Therefore, the AC-TR algorithm minimizes high amplitude peaks resulting in PAPR CCDF close to the result of the PAPR-TR algorithm. 
The comparison of Fig. \ref{fig_SDR_IBO} and Fig. \ref{fig_PAPR} shows that the minimization of the PAPR metric is an efficient approach to nonlinear distortion minimization only for relatively high IBO and high $p$ values, i.e., soft limiter-like PA. 

\begin{figure}[!t]
\centering
\includegraphics[width=3.5in]{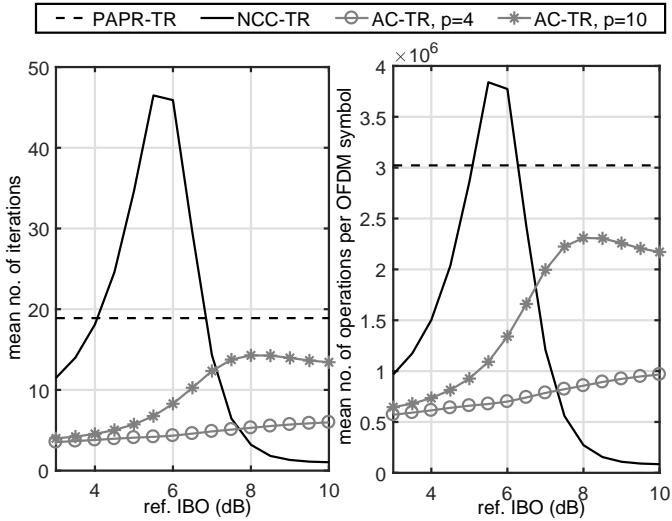}
\caption{Mean number of iterations and operations required by a given TR algorithm with respect to reference IBO and $p$ value. Stopping criteria: maximal absolute value of TR symbol improvement below 0.01.}
\label{fig_iterations}
\end{figure}

While, according to Table \ref{table_complexity}, the AC-TR and PAPR-TR algorithms have similar computational complexity, the NCC-TR algorithm typically requires fewer operations in a single iteration. However, the convergence of all algorithms can be different. As  all algorithms have the same stopping criterion, i.e., the maximal absolute value of TR symbol improvement below 0.01, fair comparison was possible. The mean number of iterations and operations per single OFDM symbol is depicted in Fig. \ref{fig_iterations}. On the left hand side, it is visible that the number of iterations required by the AC-TR algorithm is typically lower than in the case of both its competitors. Even though the NCC-TR algorithm requires fewer iterations when IBO is relatively high (above 7 dB), it is obtained at the cost of weaker nonlinear distortion power reduction abilities, as depicted in Fig. \ref{fig_SDR_IBO}. The same conclusions can be drawn from the right plot. Although the computational complexity per single iteration is the highest and mean number of operations rises with the increasing $p$ parameter, the AC-TR algorithm requires the lowest mean number of operations per OFDM symbol in all useful algorithm configurations, i.e., the NCC-TR algorithm does not provide significant nonlinear distortion power mitigation for high IBO values.       

\begin{figure}[!t]
\centering
\includegraphics[width=3.5in]{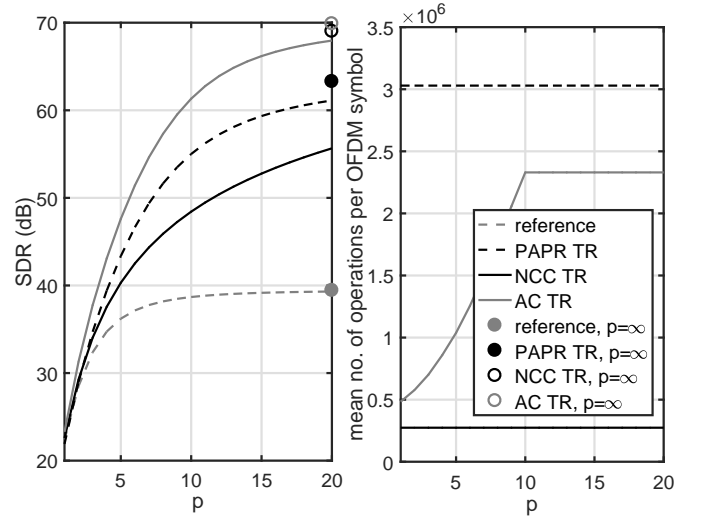}
\caption{SDR and mean number of operations as a function of $p$. Reference IBO=8 dB. For $p$ greater than 10 AC-TR uses $p=10$.}
\label{fig_SDR_vs_p}
\end{figure}

It is visible in Fig. \ref{fig_iterations} that the highest mean number of AC-TR iterations is required for IBO equal to 8 dB and $p$ equal to 10. The simulation results for a changing $p$ value for IBO equal to 8 dB is shown in Fig. \ref{fig_SDR_vs_p}. It is visible in the right-hand plot that the mean number of iterations per OFDM symbol strongly increases with $p$ for the AC-TR algorithm. As such, for $p$ values greater than $10$, $p$ equal to $10$ is used in AC-TR optimization. This helps to limit computational complexity at the cost of accuracy of PA modeling in the case of the AC-TR algorithm. However, as it is visible in the left-hand plot, the AC-TR algorithm outperforms all other algorithms in the whole range of $p$ parameters in SDR values. Even for $p=\infty$, i.e., soft-limiter PA, AC-TR outperforms the NCC-TR algorithm by about $0.86$ dB. It is significant, as AC-TR operates with an inaccurate PA model while NCC-TR is designed for soft-limiter PA. 
\subsubsection{Hardware validation}
\begin{figure}[!t]
\centering
\includegraphics[width=3.5in]{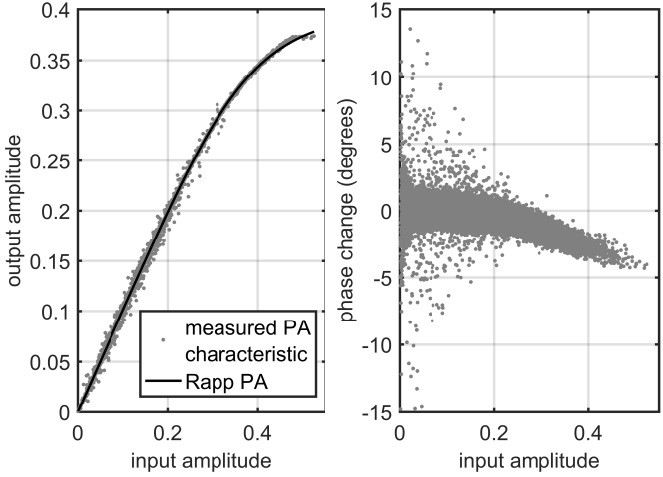}
\caption{AM-AM and AM-PM characteristic of the measures PA and Rapp model fit.}
\label{fig_AMAM_AMPM}
\end{figure}

\begin{figure}[!t]
\centering
\includegraphics[width=3.5in]{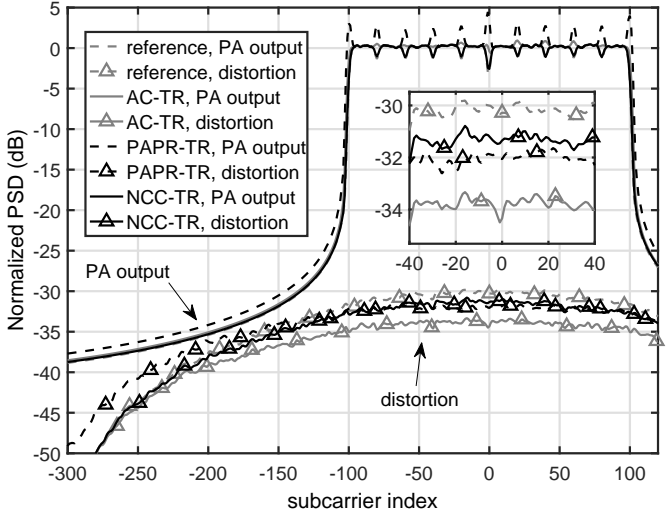}
\caption{Normalized PSDs in real transmission for reference IBO=7.8 dB and $p=2.25$.}
\label{fig_PSD_meas}
\end{figure}

Finally, the usefulness of the AC-TR algorithm in the minimization of nonlinear distortion power is shown in a Software Defined Radio (SDR) platform. The transmitter composes of a GNU Radio SDR connected with an Ethernet cable to an USRP N200 frontend. The utilized daughterboard is WBX tuned to 500 MHz carrier frequency, sampling rate 12.5 Msps and maximal TX gain. In order to minimize the effects of frontend imperfections other than nonlinearity, e.g., local oscillator leakage, IQ imbalance, phase noise, a number of measures were taken. First, an 10 MHz frequency offset was generated in USRP FPGA moving the oscillator leakage and signal mirror out of the observed band. Additionally, a GPS disciplined oscillator is used in order to minimize oscillator phase noise. The receiver is build using a Rohde\&Schwarz FSL6 spectrum analyzer. It is connected to a laptop using the Matlab Instrument Control Toolbox. The spectrum analyzer receives raw IQ samples for further processing. The transmitted sequence consists of 100 random OFDM symbols using the same parameters as in the case of simulations. The procedure for PA characteristic extraction used in this setup is given in \cite{Kryszkiewicz_ISWCS_2015_predistortion}. The resultant AM-AM and AM-PM characteristics, as well as the least squares Rapp model fit with $p=2.25$ and IBO equal to 7.8 dB are shown in Fig. \ref{fig_AMAM_AMPM}. The non-zero AM-PM characteristic reveals a memory effect featuring the measured PA. Although the Rapp model does not model this effect, it is shown to be accurate enough to make the AC-TR method outperform all other compared methods in nonlinearity distortion minimization. The resultant PSD plots are shown in Fig. \ref{fig_PSD_meas}. While high sidelobes power does not allow us to see the performance improvement for the whole signal observed at the frontend output, the distortion signal shows significant differences. The AC-TR achieves the lowest distortion power. It is most visible in-band of the OFDM signal where the gain in comparison to the reference system equals about 4 dB. As for the SDR value reference, PAPR-TR, NCC-TR and AC-TR achieve 28.8 dB, 29.5 dB, 29.7 dB and 31.7 dB, respectively. These SDR results are coherent with simulation results for the same set of parameters $p$ and IBO. NCC-TR, PAPR-TR and reference systems achieve SDR higher by about 0.45 dB in simulations as a result of some distortions introduced by practical TX/RX frontend or signal processing, e.g., synchronization. In the case of AC-TR, the gain of simulations is 0.7 dB. It is caused both by the above mentioned imperfection and the inaccuracy in modeling of practical nonlinearity by the Rapp model.  

\section{Conclusion}
 The utilization of the knowledge on effective PA characteristics obtained by digital predistortion methods can significantly improve TR algorithm efficiency in terms of the resultant SDR. The proposed AC-TR algorithm provides the highest SDR in comparison to two other state-of-the-art TR algorithms, at a typically lower number of operations required. The proposed AC-TR method converges relatively fast to the optimal solution and the computational cost of a single iteration is acceptable thanks to (I)FFT algorithm utilization.  
\appendices
\section{Proof of Optimization Problem Convexity}
\label{sec_appendix_convex}
In order to prove the convexity of $f(\mathbf{d_{\mathrm{TR}}})$, it is enough to do it for a single $n$ sample, as the sum of convex functions is a convex function \cite{boyd2004convex}. Additionally, affine mapping of $\mathbf{d_{\mathrm{TR}}}$ to $\Re y_n$ and $\Im y_n$ does not change the convexity of the considered function \cite{boyd2004convex}. Thus, it is enough to consider function 
\begin{align}
f_n(\Re y_n,\Im y_n)=&\left( \left(\Re y_n\right)^{2}+\left(\Im y_n\right)^{2}\right)
\\ \nonumber & \!\!\!\!\!\cdot
\left(\left(1+\frac{
\left( \left(\Re y_n\right)^{2}+\left(\Im y_n\right)^{2}\right)^{p}
}{V^{2p}}\right)^{-\frac{1}{2p}}\!\!-\!\!K\right)^{2}. 
\end{align}
Next, it can be observed that $\left(\Re y_n\right)^{2}+\left(\Im y_n\right)^{2}$ is convex, so it has to be proved that function
 \begin{align}
f_n(q)=q
\left(\left(1+\frac{
q^{p}
}{V^{2p}}\right)^{-\frac{1}{2p}}\!\!-\!\!K\right)^{2} 
\end{align}
is convex and non-decreasing \cite{boyd2004convex} for $q=\left(\Re y_n\right)^{2}+\left(\Im y_n\right)^{2}$ and $q\geq 0$.  
Function $f_n(q)$ is non-decreasing if the first-order derivative is non-negative for $q\geq 0$. The derivative equals
\begin{align}
f_n^{'}(q)\!=\!
\left(\!K\!\!-\!\!
\left(1+\frac{q^{p}}{V^{2p}}\right)^{\!\!-\frac{1}{2p}}\right)
\left(\!K\!\!-\!\!
\left(1+\frac{q^{p}}{V^{2p}}\right)^{\!\!-1-\frac{1}{2p}}\right)
. 
\end{align}
As $q\geq 0$ and $V>0$, it can be shown $1+\frac{q^{p}}{V^{2p}}\in \langle 1, \infty)$. For all negative powers, i.e., $x\in(-\infty,0)$, it is obtained  $\left(1+\frac{q^{p}}{V^{2p}}\right)^{x}\in (0,1\rangle$. In such a case, both factors $K\!\!-\!\!
\left(1+\frac{q^{p}}{V^{2p}}\right)^{-\frac{1}{2p}}$ and $K\!\!-\!\!
\left(1+\frac{q^{p}}{V^{2p}}\right)^{-1-\frac{1}{2p}}$ are non-negative at least for $K\geq 1$. As a result, the first-order derivative is non-negative as well.
The necessary and sufficient convexity condition is $ f_n^{''}(q)\geq0$.
The second order derivative is
\begin{align}
f_n^{''}(q)&=\frac{pq^{p-1}}{V^{2p}}
\left(1+\frac{q^{p}}{V^{2p}}\right)^{-\frac{1}{2p}-1}
\\ \nonumber &
\cdot \Biggl(
\frac{1}{2p}
\left(K-
\left(1+\frac{q^{p}}{V^{2p}}\right)^{-\frac{1}{2p}-1}\right)
\\ \nonumber &
\!\!\!\!+\left(\frac{1}{2p}+1\right)
\left(1+\frac{q^{p}}{V^{2p}}\right)^{-1}
\left(K-
\left(1+\frac{q^{p}}{V^{2p}}\right)^{-\frac{1}{2p}}\right)
\Biggr)
. 
\end{align}
As $q\geq 0$ and $V>0$, it can be shown that $\frac{pq^{p-1}}{V^{2p}}\geq 0$. Having in mind that $\left(1+\frac{q^{p}}{V^{2p}}\right)^{x}\in (0,1\rangle$ for $x\in(-\infty,0)$, it can be stated that for $K\geq 1$, both additive components, starting with multipliers $\frac{1}{2p}$ and $\frac{1}{2p}+1$, are non-negative, and the whole function $f_n^{''}(q)$ non-negative. Therefore, function $f(\mathbf{d_{\mathrm{TR}}})$ is convex.

\section{Proof of equivalence between (\ref{eq_B}) and (\ref{eq_szybkie_B})}
\label{sec_appendix_B}
The equivalence between directly calculated Hessian matrix entries and their fast calculation strategy using the FFT block is shown here. As an example, the derivation of (\ref{eq_B}) based on (\ref{eq_szybkie_B}) is provided. By introducing (\ref{eq_DFT_fast}) into (\ref{eq_szybkie_B}), the following is obtained:
\begin{align}
B^{(k)}_{l,m}\!\!&=
\frac{0.5}{N}\Im \!\left( \sum_{n=0}^{N-1} \Lambda^{(k)}_{n}y^{(k) 2}_{n} e^{-\imath 2 \pi \frac{n (T_l+T_m)}{N}}\!\right)\!
\\ & \nonumber
-\!\frac{1}{N}\Im \! \!\left( \!\sum_{n=0}^{N-1} \!\!\left(\Gamma^{(k)}_{n}\!\!+\!0.5\Lambda^{(k)}_{n}|y^{(k)}_{n}|^{2}\right) \!e^{-\imath 2 \pi \frac{n (T_l-T_m)}{N}}\!\right)
\\ & \nonumber
=\frac{1}{N}\sum_{n=0}^{N-1} 0.5\Lambda^{(k)}_{n}\Im \left(y^{(k) 2}_{n} e^{-\imath 2 \pi \frac{n T_l}{N}}e^{-\imath 2 \pi \frac{n T_m}{N}}\right)
\\ & \nonumber
-\left(\Gamma^{(k)}_{n}+0.5\Lambda^{(k)}_{n}|y^{(k)}_{n}|^{2}\right)
\Im\left(e^{-\imath 2 \pi \frac{n T_l}{N}}e^{\imath 2 \pi \frac{n T_m}{N}}\right).
\end{align}
Using the same substitution as used in (\ref{eq_time_domain_simplified}) that 
$F_{n,l}=\frac{1}{\sqrt{N}}e^{-\imath 2 \pi \frac{nT_l}{N}}$ and after some multiplications it is obtained
\begin{align}
\nonumber
\!\!&B^{(k)}_{l,m}\!\!=\!\!\!
\nonumber
\sum_{n=0}^{N-1} \!\!0.5\Lambda^{\!\!(k)}_{n}\!\Bigl(\!
\left(\!\Re y_n^{(k)2}\!\!-\!\!\Im y_n^{(k)2}\!\right)\!
\left(\!\Re F_{\!n,l}\Im F_{\!n,m}\!\!+\!\!\Im F_{\!n,l}\Re F_{\!n,m}\!\right)
\\ &
+\!2\Re y_n^{(k)}\Im y_n^{(k)}\Re F_{\!n,l}\Re F_{\!n,m}
\!\!-\!\!2\Re y_n^{(k)}\Im y_n^{(k)}\Im F_{\!n,l}\Im F_{\!n,m}\!\!\Bigr)\!
\\ & \nonumber
-\!\left(\Gamma^{(k)}_{n}\!+\!0.5\Lambda^{(k)}_{n}|y^{(k)}_{n}|^{2}\right)
\left(\!-\Re F_{n,l}\Im F_{n,m}\!+\!\Im F_{n,l}\Re F_{n,m}\right).
\end{align}
After further multiplications and addition/subtraction of repeating components it is obtained
\begin{align}
B^{(k)}_{l,m}\!\!&=
\sum_{n=0}^{N-1} 
\Gamma^{(k)}_{n} \left(
\Re F_{n,l}\Im F_{n,m}-\Im F_{n,l}\Re F_{n,m}
\right)
\\ & \nonumber
+\Lambda^{(k)}_{n}\Bigl(
\Re y_n^{(k)2}\Re F_{n,l}\Im F_{n,m}
-\Im y_n^{(k)2}\Im F_{n,l}\Re F_{n,m}
\\ &
+\Re y_n^{(k)}\Im y_n^{(k)}\Re F_{n,l}\Re F_{n,m}
-\Re y_n^{(k)}\Im y_n^{(k)}\Im F_{n,l}\Im F_{n,m}
\Bigr),
\nonumber
\end{align}
that can be simply converted to (\ref{eq_B}).

\ifCLASSOPTIONcaptionsoff
  \newpage
\fi

\bibliographystyle{IEEEtran}
\bibliography{pawla_bib}


\end{document}